\PassOptionsToPackage{svgnames}{xcolor}
\documentclass{vgtc}                          % final (conference style)
\graphicspath{{figures/}{pictures/}{images/}{./}} % where to search for the images

\usepackage{times}                     % we use Times as the main font
\usepackage{amssymb}
\usepackage{tabu}                      % only used for the table example
\usepackage{booktabs}                  % only used for the table example
\usepackage{lipsum}                    % used to generate placeholder text
\usepackage{mwe}                       % used to generate placeholder figures
\usepackage{amsmath}                    % used to generate equation
\usepackage{mathptmx}                  % use matching math font
\usepackage{graphicx} %use graph format
\usepackage{epstopdf}
\usepackage{svg}
\usepackage{multirow}
\usepackage [english]{babel}
\usepackage [autostyle, english = american]{csquotes}
 \usepackage[final]{changes} %不带标记版本

\definechangesauthor[name={adder}, color=blue]{+} %增加标注
\definechangesauthor[name={deleter}, color=red]{-} %删除标注
\newcommand{\ADD}[1]{\added[id={+}]{#1}}
\newcommand{\DEL}[1]{\deleted[id={-}]{#1}}

\MakeOuterQuote{"}

\onlineid{1032}

\vgtccategory{Research}

\vgtcinsertpkg

\title{Impact of \DEL{Object} \ADD{Target} and Tool Visualization on Depth Perception and Usability in Optical See-Through AR}

\author{
Yue Yang\thanks{first author, e-mail: yueyang1@stanford.edu.}\\ Stanford University%
        \and Xue Xie \\SJTU
\and Xinkai Wang \\SEU
\and Hui Zhang \\HUST
\and Chiming Yu \\HUST
\and Xiaoxian Xiong \\SJTU
\and Lifeng Zhu \\ %
     SEU %
    \and Yuanyi Zheng \\SJTU
\and Jue Cen \\SJTU
\and Bruce Daniel \\ Stanford University%
\and Fred Baik \thanks{corresponding author, e-mail: fbaik@stanford.edu.}\\ %
     Stanford University %
     }

\teaser{
  \centering
  \includegraphics[width=\linewidth]{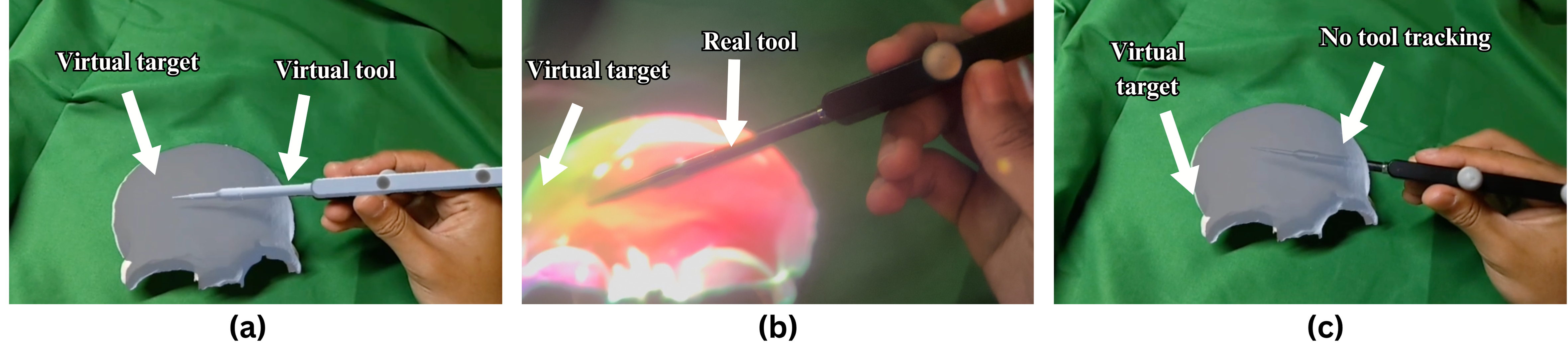}
  \caption{Three visualization conditions in HoloLens 2: (a) Virtual Tool (T1) – opaque holographic tool at tracked pose; (b) Real Tool (T2) – physical tool shown via occlusion masking; (c) No Tool (T3) – tool untracked, virtual target always on top.}
  \label{fig:1}
}

\abstract{
     Optical see-through augmented reality (OST-AR) systems like Microsoft HoloLens 2 hold promise for arm's distance guidance (e.g., surgery), but depth perception of the hologram and occlusion of real instruments remain challenging. We present an evaluation of how visualizing the target object with different transparencies and visualizing a tracked tool (virtual proxy vs. real tool vs. no tool tracking) affect depth perception and system usability. 10 participants performed two experiments on HoloLens 2. In Experiment 1, we compared high-transparency vs. low-transparency \ADD{target} \DEL{object} rendering in a depth matching task at arm’s length. In Experiment 2, participants performed a simulated surgical pinpoint task on a frontal bone target under six visualization conditions (2 \ADD{target} \DEL{object} transparencies × 3 tool visualization modes: virtual tool hologram, real tool, or no tool tracking). We collected data on depth matching error, target localization error, system usability, task workload, and qualitative feedback. Results show that a more opaque target yields significantly lower depth estimation error than a highly transparent target at arm's distance. Moreover, showing the real tool (occluding the virtual target) led to the highest accuracy and usability with the lowest workload, while not tracking the tool yielded the worst performance and user ratings. However, making the target highly transparent, while allowing the real tool to remain visible, slightly impaired depth cues and did not improve usability. Our findings underscore that correct occlusion cues – rendering virtual content opaque and occluding it with real tools in real-time – are critical for depth perception and precision in OST-AR. Designers of arm-distance AR systems should prioritize robust tool tracking and occlusion handling; if unavailable, cautiously use transparency to balance depth perception and tool visibility.
} % end of abstract


\begin{document}


\firstsection{Introduction}
\maketitle
Augmented reality (AR) is emerging as a tool for surgical navigation, overlaying digital anatomical models and procedural cues directly onto patients. Optical see-through head-mounted displays (OST-HMDs), such as the Microsoft HoloLens 2, allow clinicians to view virtual holograms within their real environment, offering potential advantages over screen-based navigation by reducing cognitive load and improving hand–eye coordination. However, clinical adoption remains limited due to perceptual, usability, and human factors challenges, particularly depth misperception and occlusion ambiguities \cite{birlo2022utility, doughty2022augmenting}. Accurate depth perception is critical, yet current OST-HMDs often distort distances and render holograms semi-transparently, undermining depth cues like occlusion \cite{adams2022depth, liu2024depth}.

Researchers have explored visual design strategies to address these issues. Enhancements include shadows, contextual references, and stronger occlusion rendering \cite{krajancich2020factored}. Another challenge is visualizing tracked surgical tools at arm’s length. Strategies include rendering virtual proxies or revealing the real tool via occlusion masking \cite{sanches2012mutual, macedo2021occlusion, tang2020grabar}, each with trade-offs in realism and control over appearance. While qualitative work suggests realistic occlusion aids spatial accuracy \cite{li2021effects}, empirical comparisons remain scarce—particularly when tools and targets overlap in surgical working distances.

This study experimentally evaluates visualization strategies for targets and tools in OST-AR using the HoloLens 2, simulating a surgical guidance task with two independent variables:  
\textbf{Target Transparency:} Virtual target rendered with either low transparency (opaque) or high transparency (see-through).  
\textbf{Tool Visualization Mode:} Tool shown as (i) virtual holographic proxy, (ii) real tool via occlusion masking, or (iii) no visual representation (untracked, no occlusion).

Thus, we test the following hypotheses:  
\textbf{H1:} Opaque targets yield more accurate depth perception than transparent ones at arm’s distance, as stronger occlusion cues aid depth judgment.  
\textbf{H2:} Task accuracy is highest when the real tool is visible and occludes the virtual target, enabling natural alignment.  
\textbf{H3:} Usability is highest and workload lowest when the real tool is visible and tracked.  
\textbf{H4:} Transparency may help when the tool is not visualized by keeping it visible through the target but may reduce depth clarity when the tool is visible.

%% \section{Introduction} %for journal use above \firstsection{..} instead
\section{Methods}
\subsection{Visualization Conditions}
All experiments used a Microsoft HoloLens 2 OST-HMD running a custom Unity 3D app, providing binocular, stereo see-through displays for real-time virtual overlays. Tool tracking followed Martin et al. \cite{martin2023sttar} using built-in sensors and passive IR markers to register the tool’s 3D pose in the HoloLens coordinate system.

Three tool visualization modes were tested:
\textbf{T1: Virtual Tool (Hologram)} — A virtual 3D model rendered at the tracked pose with an opaque shader, allowing occlusion of the virtual target for spatial coherence.  
\textbf{T2: Real Tool (See-through Occlusion)} — The physical tool shown directly; tracked pose masked virtual content to preserve natural occlusion.  
\textbf{T3: No Tool Tracking} — Tool untracked and unrendered; virtual target always visible, breaking occlusion.

The virtual target, a frontal skull bone model, appeared in two styles:  
\textbf{O1: High Transparency} — ~50\% opacity via Unity’s standard shader (alpha 0.5), further lightened by OST-HMD additive blending, enabling an “X-ray” view.  
\textbf{O2: Low Transparency (Opaque)} — Default material for maximum perceived solidity.

Experiment 1 compared O1 vs. O2 for perceived depth at arm’s distance (H1). Experiment 2 used a 2×3 within-subjects design crossing target transparency (high/low) with tool mode (T1–T3) to test H2–H4 (\cref{fig:1}).  

Ten participants (4 male, 6 female; age 21–30) with normal/ corrected vision and limited AR/VR experience completed two tasks with breaks to reduce fatigue, after giving informed consent.

\subsection{Depth Perception Task}
\begin{figure}[tbh]
\begin{centering}
    \includegraphics[scale=0.3]{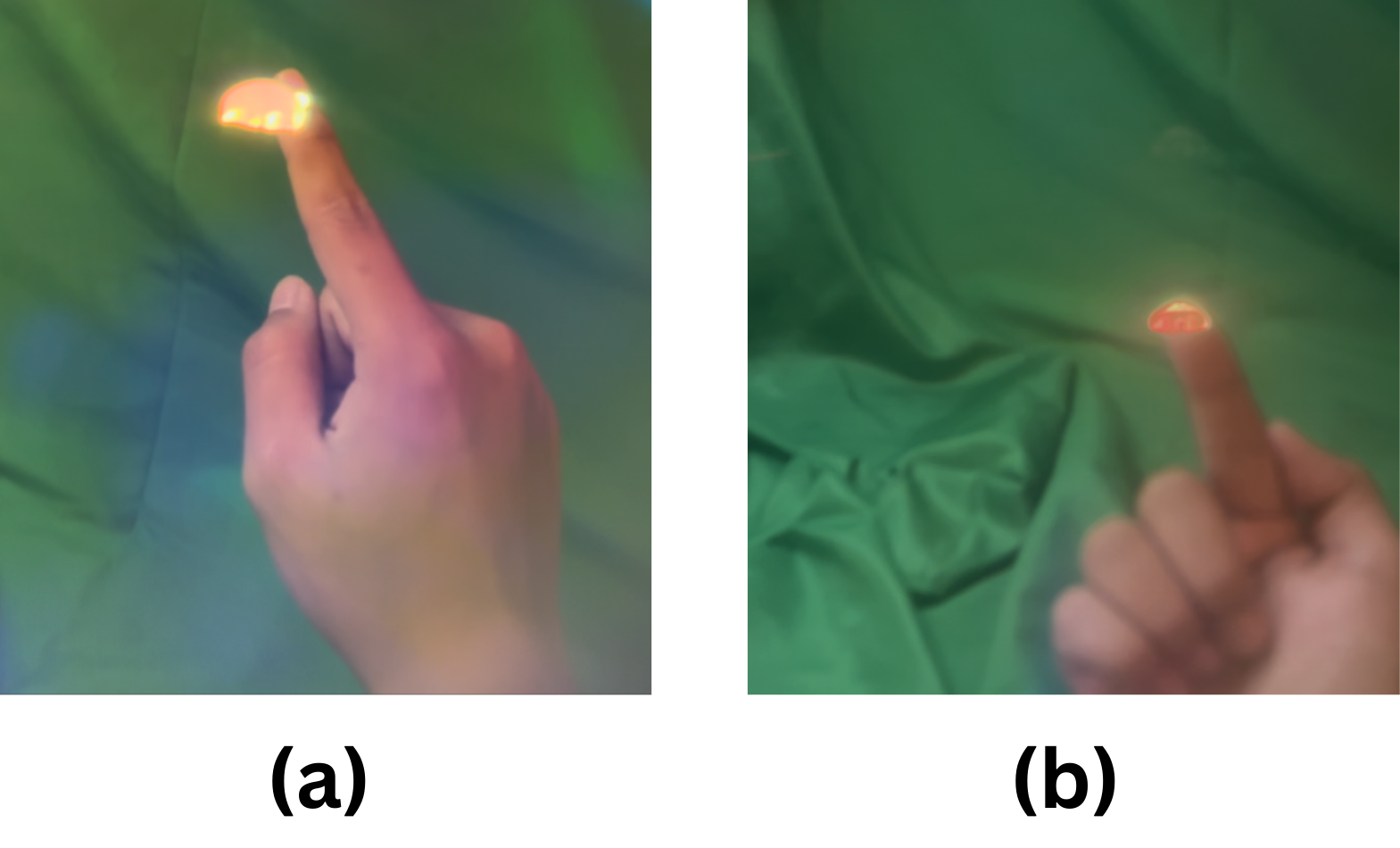}
    \caption{Two visualizations of the depth perception task, captured through first-person point of view through Hololens 2 lenses. Participants' hand tracking data are streamlined through HL2SS to a workstation for post-experiment analysis. }
    \label{fig:3}
    \end{centering}
\end{figure}
The first experiment evaluated how the transparency of a virtual target affects depth perception in OST-AR. Participants performed a depth-matching task at approximately arm’s length, using their index finger to physically reach toward the perceived location of a virtual \ADD{target} \DEL{object} rendered in the Microsoft HoloLens 2.

In each trial, a virtual \ADD{target} sphere (2 cm diameter) was rendered at a fixed distance in front of the user, appearing at a randomly selected position following the headset's head pose at a relative depth between 25 and 40 \DEL{mm} \ADD{cm}. Participants were instructed to reach out and touch the perceived center of the \DEL{virtual sphere } \ADD{target} using their dominant index finger, without any feedback or visual confirmation of contact. Participants viewed the scene binocularly through the headset and were allowed to freely move their hands and disencouraged from head movement. Once they believed their fingertip was aligned with the center of the virtual sphere in depth, the position of their fingertip relative to Hololens world coordinate was recorded using the HoloLens hand-tracking system through HL2SS package \cite{hl2ss}. Specifically, we measured the Euclidean distance between the center coordinate of the virtually rendered \ADD{target} \DEL{object} and the finger tip coordinate through hand-tracking, assigning it a negative value if hand position lies between head pose and virtual \ADD{target} \DEL{object}. 

We tested two rendering conditions in a within-subject design: high transparency (O1) and low transparency (O2). In the O1 condition, the virtual sphere appeared as a translucent ghost—participants could see the background and their own hand through the \ADD{target} \DEL{object}. In contrast, the O2 condition rendered the sphere as a more opaque solid. Participants were not informed of the differences between conditions and were simply asked to perform the task as accurately as possible. Each participant completed 10 trials under each transparency condition. Prior to data collection, participants completed a brief practice block \ADD{using both transparency conditions} to familiarize themselves with the task and the AR environment. 

The primary outcome measure was signed depth error, computed as the difference between the participant’s perceived depth and the true depth of the virtual sphere. Positive errors indicated overshooting—placing the finger beyond the virtual target—while negative errors indicated undershooting. We also analyzed the absolute (unsigned) error to assess overall precision. We hypothesized (H1) that participants would achieve more accurate depth matches (i.e., smaller mean signed error) when the sphere was rendered opaquely, due to the presence of stronger occlusion cues and improved visual solidity. Other conditions (color, texture, etc.) are controlled. 

\subsection{Tool Interaction Task}
The second experiment examined how target transparency and tool visualization jointly affect performance and usability in a simulated surgical localization task. Participants used a real handheld tool to perform a precision “pinpointing” task on a virtual human skull model, simulating frontal bone localization during craniofacial surgery.

A 3D skull model was rigidly registered to AR space and aligned with a mannequin head using a custom ICP-based algorithm on the HoloLens 2 depth sensor. The target region was the forehead. In each trial, a 2 mm red dot marked a predefined surface location, and participants aligned the tool tip with this target in 3D space. The setup allowed simultaneous viewing of the virtual skull, the underlying 3D-printed skull, and the real tool.
\begin{figure}[tbh]
\begin{centering}
    \includegraphics[width = \linewidth]{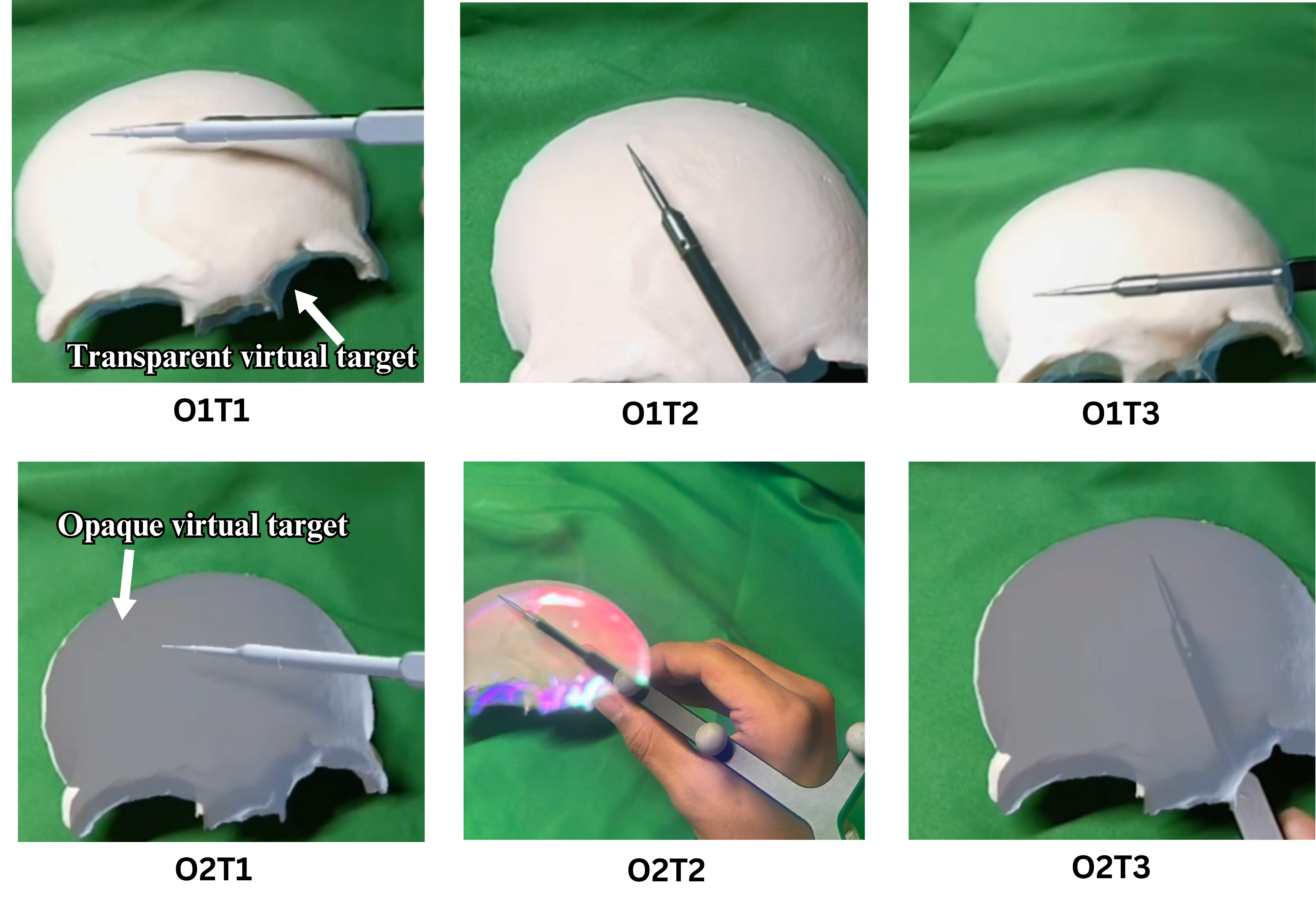}
    \caption{Six combinations of target transparency (O1: high, O2: low) and tool visualization (T1: virtual, T2: real with occlusion masking, T3: none). Due to the additive nature of OST-HMDs and capture limitations, O1 appears less distinct in recordings, with O1T2 and O1T3 looking similar despite perceptual differences through the HoloLens. For O2T3, a through-the-lens capture better represents the user’s view.}
    \label{fig:4}
    \end{centering}
\end{figure}

Visualization conditions followed a 2 (Target Transparency: O1 high vs. O2 low) × 3 (Tool Visualization: T1 virtual tool, T2 real tool with occlusion masking, T3 no tool) within-subject design (\cref{fig:4}). The tool was an Optitrack digitizing probe (270 mm) with four IR markers, pivot-calibrated before trials \cite{optitrack_2025}.

Each participant completed 8 trials per condition, with 10 unique dot locations distributed across the frontal bone. Condition and target orders were randomized. The tool tip’s 3D position was logged on verbal cue, and localization error (mm) was computed as the Euclidean distance to the ground truth. Median error and IQR per condition were used for statistical analysis.

After each block, participants completed the NASA-TLX \cite{hart2006nasa} for workload and the SUS \cite{bangor2008empirical} for usability (0–100 scale). Post-experiment interviews captured preferences and perceived limitations. In total, 80 localization data points per participant were collected, addressing H2 and H3 by evaluating accuracy, workload, and usability across AR visualization modes.

\begin{figure*}[tbh]
\begin{centering}
    \includegraphics[width = \linewidth]{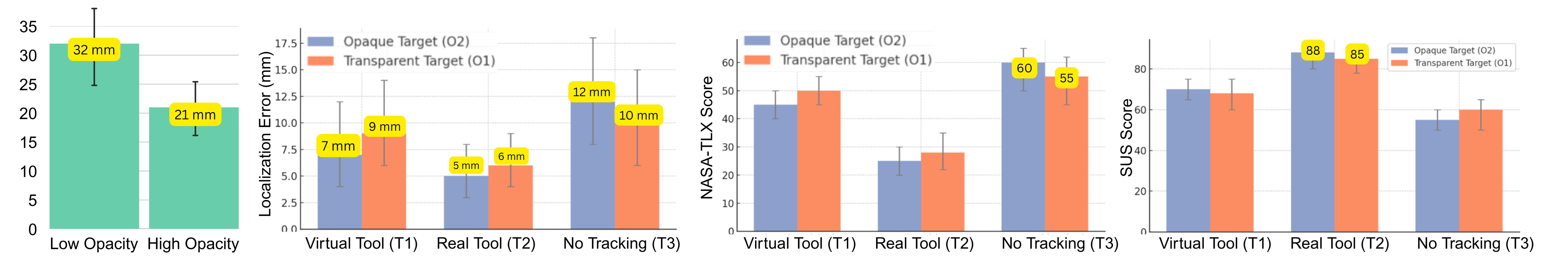}
    \caption{Quantitative evaluation of experimental results. Data are reported in median with IQR bar. Critical values are highlighted and labeled. }
    \label{fig:6}
    \end{centering}
\end{figure*}
\section{Results}
\subsection{Depth Perception Task (H1)}
In the depth matching task with 10 participants, we observed a strong effect of \ADD{target} \DEL{object} transparency on perceived depth. When the virtual \ADD{target} \DEL{object} was rendered with high transparency (O1), participants consistently placed it farther than its actual depth, resulting in large positive signed errors. In contrast, the low-transparency (more opaque) \ADD{target} \DEL{object} (O2) led to smaller errors that were more evenly distributed around zero.

\cref{fig:6} summarizes the results: the median signed error for the transparent condition was +32 mm \ADD{(IQR 25-38 mm)}, while the opaque condition yielded a smaller median error of +21 mm \ADD{(IQR 16-25.5 mm)}. This difference in depth judgment was statistically significant (p $<$ 0.05) based on a Wilcoxon signed-rank test following non-normality results. Thus, participants were much more accurate in depth matching with an opaque \ADD{target} \DEL{object}, suggesting that increased opacity improved accuracy. The effect was also directionally consistent—participants overshot the transparent \ADD{target} \DEL{object} more often, likely due to depth cue conflicts caused by seeing the reference \ADD{target} \DEL{object} through the ghost-like hologram. These results support our hypothesis (H1): reducing transparency and enhancing occlusion cues significantly improves depth perception accuracy in AR.

\subsection{Tool Interaction Task (H2, H3, H4)}
\subsubsection{Localization Accuracy}
The pinpointing task results show strong main effects of the tool visualization method on accuracy. \cref{fig:6} presents the median localization error for each condition. Overall, conditions with no tool tracking (T3) had the largest errors, while conditions with the real tool visible (T2) had the smallest errors. For example, with an opaque target, median error was 12 mm \ADD{(IQR 8-18 mm)} in O2\DEL{+}T3 (no tool) versus 5 mm \ADD{(IQR 3-8 mm)} in O2\DEL{+}T2 (real tool). The virtual tool (T1) was intermediate (7 mm \ADD{(IQR 4-12 mm)} in O2\DEL{+}T1). Transparent target conditions followed a similar pattern: O1\DEL{+}T3 median 10 mm \ADD{(IQR 6-15 mm)}, O1\DEL{+}T2 6 mm \ADD{(IQR 4-9 mm)}, O1\DEL{+}T1 9 mm \ADD{(IQR 6-14 mm)}. We note that making the target transparent did not uniformly improve accuracy—in fact, when the tool was tracked (T1 or T2), transparency slightly increased error (compare 9 mm vs 7 mm for virtual tool; 6 mm vs 5 mm for real tool). However, for the no-tracking case, transparency (O1) reduced error from 12 mm to 10 mm. This suggests a small interaction: transparency helped only when the system lacked proper tool occlusion (since seeing the tool through the target aided the user), but otherwise, transparency was detrimental for depth/precision.

Experiment 2’s objective results demonstrate that not visualizing the tool incurs a large performance penalty. Participants struggled when the virtual target did not get occluded by the real scalpel, often overshooting or misaligning because the target graphic would obscure where their scalpel tip really was. Many participants in debriefing mentioned this was “like trying to touch something with your hand behind a curtain”. Conversely, both methods of showing the tool (virtual or real) allowed much more precise pointing, roughly doubling accuracy (errors 5–9 mm vs up to 15 mm in worst cases). The best performance occurred when the \ADD{real} \DEL{actual} tool was visible and properly occluding the target, closely simulating real-world direct vision. Interestingly, the virtual tool hologram was almost as effective, indicating AR proxies can work if well calibrated. The effect of target transparency on accuracy was secondary to the tool effect, though we did observe the hypothesized slight trade-off: transparency helps when occlusion is absent, but otherwise an opaque target was optimal for accuracy.

\subsubsection{Subjective Workload}
The self-reported NASA-TLX workload scores mirrored the objective difficulty of each condition. \cref{fig:6} summarizes the median TLX results for each visualization method. Lower scores indicate lower perceived workload. Participants reported the highest workload in the no tool conditions (T3), especially with an opaque target (which was described as most frustrating). The median TLX overall score for O2\DEL{+}T3 was 60 (IQR 50–65) on the 0–100 scale, and for O1\DEL{+}T3 was 55 (IQR 45–62). In contrast, conditions with the real tool (T2) had much lower workload: median 25 for O2\DEL{+}T2 (IQR 20–30) and 28 \ADD{(IQR 22-35)} for O1\DEL{+}T2. The virtual tool (T1) was intermediate: median TLX increased from 45 \ADD{(IQR 40-50)} to 50 \ADD{(IQR 45-56)} when changing from O2 to O1. In other words, having no tool visualization roughly doubled the perceived workload compared to having the real tool visible. Participants specifically cited mental demand and frustration as much higher in T3 – they had to “guess where the tip was” and felt uncertain, which was stressful. A Friedman test on TLX scores across the six conditions was significant (p $<$ 0.001). Post-hoc comparisons showed T3 conditions had significantly higher workload than T1 or T2 (p $<$ 0.001 in all cases). For example, O2\DEL{+}T3 vs O2\DEL{+}T2: all participants reported higher workload in the former (by 20–40 points). T1 also had higher workload than T2 (p $<$ 0.01): even though performance was similar, participants felt using the holographic tool required more concentration. They reported that the virtual tool’s appearance sometimes lagged or was slightly misaligned, requiring conscious effort to use, whereas the real tool felt natural. Transparency had no significant main effect on TLX; differences of 3–5 points were observed (transparent target caused slightly higher mental demand in T1/T2, but slightly less in T3), but these canceled out overall. Many participants didn’t explicitly notice the transparency change affecting workload, saying they “adapted quickly” to either view.

\subsubsection{System Usability}
The System Usability Scale (SUS) results reflected a similar pattern regarding overall user satisfaction with each condition (see \cref{fig:6}). Higher SUS means better perceived usability. Participants rated the AR system with real tool visualization (T2) most favorably, with median SUS scores of 88 \ADD{(IQR 80-92)} for O2 and 85 \ADD{(IQR 78-89)} for O1, which are in the “excellent” range \ADD{\cite{bangor2009determining}}. The virtual tool condition (T1) was rated moderately usable: SUS = 70 \ADD{(IQR 65-75)} for O2 and 68 for O1 \ADD{(IQR 60-75)}, around the threshold of acceptable. The no-tool condition received poor usability scores: SUS 55 \ADD{(IQR 51-60)} for O2 and 60 \ADD{(IQR 50-64)} for O1, which are considered “below average” (SUS $<$ 68 is sub-average). A Friedman test on SUS was significant (p $<$ 0.001). Pairwise tests showed all differences among T3 vs T1 vs T2 were significant (p $<$ 0.05). In line with H3, users greatly preferred having the tool visible. The difference between virtual and real tool was also notable, indicating a genuine usability benefit to seeing the \ADD{real} \DEL{actual} instrument. Users preferred the virtual tool to having nothing: one user said, “Using the ghost scalpel (virtual) was slightly weird at first but at least I knew where my tool was. The worst was when the hologram [target] blocked my scalpel completely. That was very hard.” The effect of transparency on SUS was small. Thus, transparency did not strongly affect perceived usability in either direction.

The subjective data strongly reinforce the objective findings: not representing the tool made the task feel difficult and frustrating, while showing the tool (especially the real one) made it easy and satisfactory. All 10 participants in their post-study interview chose the real tool + opaque target (O2\DEL{+}T2) as their preferred method, often citing it “felt the most real and precise.” The virtual tool was the second choice for most, with comments like “it was pretty good once I got used to it, but I occasionally doubted if the hologram tip exactly matched my real tip.” The no-tool condition was nobody’s preference; it was universally described with negative terms (frustrating, confusing, clumsy). As for target transparency, opinions were split: a slight majority (8/10) said they preferred the opaque target because it “looked more solid and gave a better sense of depth,” supporting our performance data. One participant insightfully said, “If I didn’t have tool tracking, then yeah, making the thing transparent helped because I could see my scalpel. But if I do have tracking, I’d rather the [virtual] bone be solid. It’s just easier to gauge where it is.”

\section{Discussion}
Our findings reinforce the critical role of occlusion cues and accurate tool visualization in improving depth perception and task performance in OST-AR. The results from Experiment 1 showed a significant improvement in depth matching accuracy when the virtual target was rendered opaquely rather than transparently. While the HoloLens 2 cannot achieve perfect opacity, even modest reductions in transparency substantially enhanced perceptual accuracy.

In Experiment 2, the real tool visualization via occlusion masking yielded the highest localization accuracy, lowest workload, and best usability scores. This suggests that enabling users to see the \ADD{real} \DEL{actual} tool in real-time, especially as it occludes virtual targets, supports natural hand–eye coordination and spatial judgment. Notably, the virtual tool proxy also performed well, demonstrating that a well-aligned holographic representation can substitute effectively when physical tool occlusion is impractical.

Interestingly, transparency showed a nuanced interaction with tool visualization: when the tool was not tracked, making the target transparent slightly mitigated performance loss by allowing users to visually locate their tool. However, in conditions where the tool was tracked, transparency degraded performance by weakening depth cues. This highlights a key trade-off—transparency may help preserve real-world visibility but can impair spatial clarity.

\section{Conclusion}
This study demonstrates that in OST-AR systems like the HoloLens 2, enhancing virtual \ADD{target} \DEL{object} opacity and supporting real tool occlusion are essential for precise depth perception and effective interaction. Designers should prioritize these features when developing AR interfaces for close-range tasks such as surgical guidance. When tool tracking is unavailable, transparency may offer a temporary workaround, but should be applied cautiously. Future work should explore adaptive rendering strategies that dynamically adjust \ADD{target} \DEL{object} transparency based on context, and investigate perceptual effects across broader user populations and task types.

\bibliographystyle{abbrv-doi-hyperref-narrow}

\bibliography{template}
\clearpage
\onecolumn
\listofchanges
\clearpage
\end{document}